\documentclass[12pt]{article}
\usepackage[hypertex]{hyperref}
\usepackage[dvipdfmx]{graphicx} 
\usepackage{graphicx,amsmath,amssymb,amsfonts,cite,bm}

\def\a{\alpha}    \def\d{\delta} \def\D{\Delta} \def\e{\epsilon}  \def\h{\eta}     \def\L{\Lambda} \def\m{\mu} \def\n{\nu}       \def\S{\Sigma} \def\t{\tau}     

\def\dg{\dagger}  

\newcommand{\diag}[2]{ \begin{pmatrix}  #1 & 0 \\ 0 & #2 \\   \end{pmatrix}  }

\newcommand{\llp}{ \left [ }
\newcommand{\rlp}{ \right ] }
\newcommand{\Lg}{\mathcal{L}}

\newcommand{\vev}[1]{ \langle {#1} \rangle }
\newcommand{\meV}{ {\rm meV} }
\newcommand{\eV}{ {\rm eV} }

\newcommand{\GeV}{ {\rm GeV} }

    \def\ds{\displaystyle}

\begin{document}

\begin{titlepage}

\begin{flushright}
STUPP-16-228
\end{flushright}

\vskip 1.35cm

\begin{center}
{\large \bf Analysis of right-handed Majorana neutrino mass \\ in an $SU(4) \times SU(2)_{L} \times SU(2)_{R}$ Pati--Salam model \\ with democratic texture}

\vskip 1.2cm

Masaki J. S. Yang

\vskip 0.4cm

{\it Department of Physics, Saitama University, \\
Shimo-okubo, Sakura-ku, Saitama, 338-8570, Japan\\
}


\begin{abstract} 

In this paper, we attempt to build a unified model with the democratic texture, 
that has some unification between up-type Yukawa interactions $Y_{\n}$ and $Y_{u}$.
Since the $S_{3L} \times S_{3R}$ flavor symmetry is chiral, 
the unified gauge group is assumed to be Pati-Salam type $SU(4)_{c} \times SU(2)_{L} \times SU(2)_{R}$.
The breaking scheme of the flavor symmetry is considered to be
 $S_{3L} \times S_{3R} \to S_{2L} \times S_{2R} \to 0$. 
In this picture, the four-zero texture is desirable for realistic masses and mixings. 
This texture is realized by a specific representation for the second breaking of the $S_{3L} \times S_{3R}$ flavor symmetry. 

Assuming only renormalizable Yukawa interactions, type-I seesaw mechanism, 
and neglecting $CP$ phases for simplicity, the right-handed neutrino mass matrix 
$M_{R}$ can be reconstructed from low energy input values.
Numerical analysis shows that 
the texture of $M_{R}$ basically behaves 
like the ``waterfall texture''.
Since $M_{R}$ tends to be the ``cascade texture'' in the democratic texture approach, 
a model with type-I seesaw and up-type Yukawa unification $Y_{\n} \simeq Y_{u}$
 basically requires fine-tunings between parameters. 
Therefore, it seems to be more realistic 
to consider universal waterfall textures for both $Y_{f}$ and $M_{R}$, 
e.g., by the radiative mass generation or the Froggatt--Nielsen mechanism. 

Moreover, analysis of eigenvalues shows that 
the lightest mass eigenvalue $M_{R1}$ is too light to achieve  
successful thermal leptogenesis.
Although the resonant leptogenesis might be possible, 
it also requires fine-tunings of parameters. 

\end{abstract} 

\end{center}
\end{titlepage}

\section{Introduction}

The flavor puzzle is one of the most stringent problems in the current particle physics. 
In particular, the fermion mixing matrices $U_{\rm CKM}$ \cite{Cabibbo:1963yz, Kobayashi:1973fv} and $U_{\rm PMNS}$ \cite{Pontecorvo:1957qd, Maki:1962mu} are curiously different.
Various models and ideas have been considered to explain the underlying flavor dynamics of the standard model (SM). 
Typical approaches treat the flavor symmetries \cite{Ishimori:2010au}, and/or specific flavor textures \cite{Fritzsch:1977vd, Fritzsch:1999ee}.
In the latter approach, many researchers have studied the democratic texture \cite{Harari:1978yi, Koide:1983qe, Koide:1989zt, Tanimoto:1989qh, Fritzsch:1989qm,Fritzsch:1995dj, Fukugita:1998vn, Tanimoto:1999pj, Haba:2000rf, Hamaguchi:2002vi,Watari:2002fd,Kakizaki:2003fc,Kobayashi:2004ha, Fritzsch:2004xc,Kobayashi:2006an, Xing:2010iu, Zhou:2011nu,Yang:2016esx}. 
In this approach, Yukawa interactions are assumed to have the ``democratic matrix'' (\ref{demo}), 
which is realized by  $S_{3L} \times S_{3R} $ symmetry. 

In order to explore a more fundamental understanding of flavor,  
building some unified model is a standard method. 
The grand unified theory (GUT) with the democratic texture is only discussed in \cite{Fukugita:1998kt,Fukugita:1999xb}, as far as the author knows. 
However, since these papers assumed a degenerated neutrino Yukawa matrix $Y_{\n}$, 
unification between $Y_{\n}$ and other $Y_{f}$ is difficult.
In this paper, we attempt to build another unified model with the democratic texture, 
which has some unification between up-type Yukawa interactions $Y_{\n}$ and $Y_{u}$.
Since the $S_{3L} \times S_{3R}$ flavor symmetry is chiral, 
the unified gauge group is assumed to be Pati--Salam (PS) type $SU(4)_{c} \times SU(2)_{L} \times SU(2)_{R}$  ($G_{422}$) \cite{Pati:1974yy}.
The breaking scheme of the flavor symmetry is considered to be
 $S_{3L} \times S_{3R} \to S_{2L} \times S_{2R} \to 0$. 
In this picture, the four-zero texture \cite{Fritzsch:1995nx,Nishiura:1999yt,Matsuda:1999yx, Xing:2003zd} is desirable for realistic masses and mixings. 
This texture is realized by a specific representation for the second breaking of the $S_{3L} \times S_{3R}$ flavor symmetry \cite{Xing:1996hi, Kang:1997uv,Mondragon:1998gy}. 

Assuming only renormalizable Yukawa interactions, type-I seesaw mechanism \cite{seesaw}, 
and neglecting $CP$ phases for simplicity, the right-handed neutrino mass matrix
$M_{R}$ can be reconstructed from low energy input values.
Numerical analysis shows that 
the texture of $M_{R}$ basically behaves  
like the ``waterfall texture'' in Table 1. 
Since $M_{R}$ tends to be the ``cascade texture'' in the democratic texture approach, 
a model with type-I seesaw and up-type Yukawa unification $Y_{\n} \simeq Y_{u}$ 
 basically requires fine-tunings between parameters (including its $CP$ phases, errors of the input parameters, and schemes of gauge symmetry breaking). 
If we realize the breaking scheme $S_{3L} \times S_{3R} \to S_{2L} \times S_{2R} \to 0$
by some mechanism, 
the sector of $\n_{R}$ might be too complicated 
to obtain cascade $Y_{f}$ and  waterfall $M_{R}$ in a unified picture. 
Therefore, it seems to be more realistic 
to consider universal waterfall textures for both $Y_{f}$ and $M_{R}$, 
e.g., by the radiative mass generation \cite{Babu:1990fr} or the Froggatt--Nielsen mechanism \cite{Froggatt:1978nt}. 
\begin{table} [ht]
\centering
\begin{math}
\begin{array}{|c|c|}
\hline
\begin{pmatrix}
\e & \e & \e \\
\e & \d & \d \\
\e & \d & 1 \\
\end{pmatrix}
&
\begin{pmatrix}
\e^{2} & \e \d & \e  \\
\e \d  & \d^{2} & \d \\
\e & \d & 1 \\
\end{pmatrix}
\\[7pt]
\hline
\text{Cascade} & \text{Waterfall} \\ \hline
\end{array}
\end{math}
\vspace{0.5cm}
\caption{The cascade and waterfall texture, with $1 \gg \d \gg \e$ \cite{Haba:2008dp}. }
\label{obs}
\end{table}

Moreover, analysis of eigenvalues shows that 
the lightest mass eigenvalue $M_{R1}$ is too light to achieve 
successful thermal leptogenesis \cite{Fukugita:1986hr}.
Although the resonant leptogenesis \cite{Pilaftsis:1997jf, Pilaftsis:2003gt} might be possible, 
it also requires fine-tunings of parameters. 

In this study, we assume only renormalizable Yukawa interactions. 
However, this strong tendency to the waterfall texture originates from 
the seesaw relation $M_{R} \sim Y_{u}^{T} Y_{u}$. 
Therefore, it would be rather robust for non-renormalizable Yukawa interactions,
as far as the type-I seesaw mechanism is assumed. 

This paper is organized as follows. 
The next section is a review of the Yukawa matrices with the democratic texture.
In Sect. 3, we construct a unified model with the $S_{3L} \times S_{3R}$ flavor symmetry.
Section 4 is a numerical analysis of mass matrix $M_{R}$ in this model.
Section 5 is devoted to conclusions.

\section{The four-zero texture from the democratic matrix approach}

The democratic matrix is defined as
\begin{align}
Y_{f}^{0} = {K_{f} \over 3} 
\begin{pmatrix} 1 & 1 & 1 \\ 1 & 1 & 1 \\ 1 & 1 & 1 \\ \end{pmatrix}
 \equiv {K_{f} \over 3} D ,
 \label{demo}
\end{align}
which is invariant under $S_{3L} \times S_{3R}$, the permutation symmetry between rows and columns.
It is diagonalized by the unitary matrix $U_{\rm DC}$
\begin{align}
U_{\rm DC} = 
\begin{pmatrix}
\frac{1}{\sqrt{2}} & \frac{1}{\sqrt{6}} &  \frac{1}{\sqrt{3}} \\
-\frac{1}{\sqrt{2}} & \frac{1}{\sqrt{6}} &  \frac{1}{\sqrt{3}} \\
0 & -\frac{\sqrt{2}}{\sqrt{3}}  & \frac{1}{\sqrt{3}} 
 \end{pmatrix} ,
\label{DMC}
\end{align}
and eigenvalues are given by $Y_{fi}^{0} = {\rm diag} (0,0, K_{f})$. 
Then, the democratic matrix produces mass only for the third generation. 
In order to provide masses for the first and second generations, 
the breaking scheme of the flavor symmetry is chosen as 
$S_{3L} \times S_{3R} \to S_{2L} \times S_{2R} \to 0$. 
Then, Yukawa matrices are represented as
\begin{align}
Y_{f} = {K_{f} \over 3} D + \d_{f} Y_{f}^{\d} + \e_{f} Y_{f}^{\e}, 
\end{align}
where $Y_{f}^{\d}, Y_{f}^{\e}$ breaks $S_{3L} \times S_{3R}$ and $S_{2L} \times S_{2R}$ respectively. 
This breaking scheme is discussed in several papers \cite{Mondragon:1998gy,Mondragon:2007af,Barranco:2010we,Canales:2013cga,Canales:2012dr,Saldana-Salazar:2015raa}. 
The origin and specific realization of this breaking scheme  
have not been discussed by the authors who proposed it. 
For example, the radiatively generated light fermion masses 
by broken $S_{3}$ symmetry \cite{Babu:1990fr} could explain this breaking scheme. 
In Ref.~\cite{Babu:1990fr}, $S_{3}$ breaking effects  
induce departures from the democratic texture only radiatively, and light fermion masses are suppressed by typical loop factors $[1 / (16 \pi^{2}) ]^{1 - 2}$. 
It naturally predicts the hierarchical relation 
\begin{align}
 K_{f} \gg \d_{f} \gg \e_{f}, 
\end{align}
which is required from realistic masses and mixings. 
A pedagogical explanation is also found in the review \cite{Babu:2009fd}.  
The following discussion is equivalent to Ref.~\cite{Mondragon:1998gy}.

The term $\d_{f} Y_{f}^{\d}$ is invariant under $S_{2L} \times S_{2R}$ between first and second indices, in order to provide mass only for the second generation. 
The most general form of the $S_{2L} \times S_{2R}$ invariant symmetric $Y_{f}^{\d}$ is
\begin{align}
Y_{f}^{\d} = %
\begin{pmatrix}
a & a & b \\
a & a & b \\
b & b & c
\end{pmatrix} .
\label{ydelta}
\end{align}
For later convenience, we parametrize $\d_f Y_{f}^{\d}$ as follows:
\begin{align}
\d_{f} Y_{f}^{\d} = \d_{f} 
\begin{pmatrix}
 \frac{\sqrt{2} r}{3}+\frac{1}{6} & \frac{\sqrt{2} r}{3}+\frac{1}{6} & -\frac{r}{3 \sqrt{2}}-\frac{1}{3} \\
 \frac{\sqrt{2} r}{3}+\frac{1}{6} & \frac{\sqrt{2} r}{3}+\frac{1}{6} & -\frac{r}{3 \sqrt{2}}-\frac{1}{3} \\
 -\frac{r}{3 \sqrt{2}}-\frac{1}{3} & -\frac{r}{3 \sqrt{2}}-\frac{1}{3} & \frac{2}{3}-\frac{2 \sqrt{2} r}{3} \\
\end{pmatrix} .
\label{2ndbreak}
\end{align} 
In Eq.~(\ref{2ndbreak}), there are only two free parameters $r, \d_{f}$. 
However, it does not lose generality, 
because one of the parameters in Eq.~(\ref{ydelta}) can be absorbed by the redefinition of $K_{f}$.
Similarly, $\e_{f} Y_{f}^{\e}$ provide mass for the first generations. 
Refs.~\cite{Kang:1997uv,Mondragon:1998gy} proposed that 
$\e_{f} Y_{f}^{\e}$ may be the doublet complex tensorial representation
of the $S_{3 (L+R)}$ diagonal subgroup:
\begin{align}
\e_{f} Y_{f}^{\e} = 
\begin{pmatrix}
\e_{1} & i \e_{2} & - \e_{1} - i \e_{2} \\
-i \e_{2} & - \e_{1} & \e_{1} + i \e_{2} \\
-\e_{1} + i \e_{2} & \e_{1} - i \e_{2} & 0
\end{pmatrix} .
\label{comptens}
\end{align}
In this case, the Yukawa matrices are approximately diagonalized as 
\begin{align}
U_{\rm DC}^{\dg} Y_{f} U_{\rm DC}^{} 
&= U_{\rm DC}^{\dg}  \left[ {K_{f} \over 3} D + \d_{f} Y_{f}^{\d} + \e_{f} Y_{f}^{\e} \right] U_{\rm DC}^{} 
=
\begin{pmatrix}
0 & \e_{f} e^{i \phi_{f}} & 0 \\
\e_{f} e^{- i \phi_{f}} & \d_{f} & r \d_{f} \\
0 & r \d_{f} & K_{f}
\end{pmatrix} ,
\label{fourzero}
\end{align}
where $\e_{f} e^{i \phi_{f}} = \sqrt{3} \, (\e_{1} + i\e_{2})$. 
Then, these Yukawa matrices lead to the ``four-zero texture'' or the ``modified Fritzsch texture'' 
\cite{Fritzsch:1995nx,Nishiura:1999yt,Matsuda:1999yx, Xing:2003zd}.
This relationship between the democratic texture and the four-zero texture is 
studied by several authors \cite{Xing:1996hi, Kang:1997uv,Mondragon:1998gy}.
In Eq.~(\ref{fourzero}), $r \sim O(1)$ is required to obtain the successful Cabibbo--Kobayashi--Maskawa (CKM) matrix. 
This is a natural condition because $S_{3L} \times S_{3R}$ breaking would produce a relation $Y_{22} \sim Y_{23}$.

For simplicity, we neglect all $CP$ phases of the Yukawa matrices (cf. $\phi_{f} = 0$ in Eq.~(\ref{fourzero}). 
The effect of $CP$ phases is discussed later. 
However, the qualitative result is considered to be rather robust with finite $CP$ phases. 

For the real Yukawa matrices, Eq.~(\ref{fourzero}) is perturbatively diagonalized as 
\begin{align}
B_{f}^{\dg} U_{\rm DC}^{\dg} Y_f U_{\rm DC} B_{f} = {\rm diag}( y_{1f}, y_{2f}, y_{3f}), 
\end{align}
where
\begin{align}
y_{1f} \simeq - {\e_{f}^{2} \over \d_{f}} - {r^{2} \e_{f}^{2} \over K_{f}}, ~~~~~~~
y_{2f} \simeq \d_{f} +  {\e_{f}^{2} \over \d_{f}} - {r^{2} \d_{f}^{2} \over K_{f}} , ~~~~~~~
y_{3f} \simeq K_f + {r^{2} \d_{f}^{2} \over K_{f}}.
\end{align}
The unitary matrix $B_f $ at leading order is found to be
\begin{align}
B_f & \simeq 
\begin{pmatrix}
1 &\ds -{\e_f \over \d_f } & 0 \\[8pt]
\ds {\e_f \over \d_f } & 1 & \ds r \, {\d_{f} \over K_{f}} \\[8pt]
\ds - r \,{\e_{f} \over K_{f}}  & \ds - r \, {  \d_{f} \over K_{f}} & 1
\end{pmatrix} 
 \simeq 
\begin{pmatrix}
1 & \ds - \sqrt{- {y_{f1} \over y_{f2}}} & 0 \\[8pt]
\ds \sqrt{- {y_{f1} \over y_{f2}}} & 1 & \ds r \, {y_{f2} \over y_{f3}} \\[10pt]
\ds -r \, \sqrt{- {y_{f1} \over y_{f2}} } \, {y_{f2} \over y_{f3}} & \ds - r \, {y_{f2} \over y_{f3}} & 1 
\end{pmatrix} .
\label{Bf}
\end{align}
Note that $y_{f1} / y_{f2} \simeq - \e_{f}^{2} / \d_{f}^{2}$ is always negative.

Therefore, the CKM matrix $V_{\rm CKM} = B_{u}^{\dg} B_{d}$ (without complex phase) is calculated as
\begin{align}
V_{\rm CKM} &\simeq 
\begin{pmatrix}
1 & \ds \sqrt{m_{u} \over m_{c}} &\ds - r \, \sqrt{m_{u} \over  m_{c} } {m_{c} \over m_{t}} \\
\ds - \sqrt{m_{u} \over m_{c}} & 1 &\ds - r \, {m_{c} \over m_{t}} \\[10pt] 
0 & \ds r \, { m_{c} \over m_{t}} & 1 
\end{pmatrix} 
\begin{pmatrix}
1 & \ds - \sqrt{m_{d} \over m_{s} } & 0 \\
 \ds \sqrt{m_{d} \over m_{s} } & 1 & \ds r \, {m_{s} \over m_{b}} \\[10pt] 
\ds - r \, \sqrt{m_{d} \over m_{s} } {m_{s} \over m_{b} } &\ds - r \, {m_{s} \over m_{b}} & 1 
\end{pmatrix} 
\\ & \simeq 
\begin{pmatrix}
1 & - \llp \sqrt{\dfrac{m_d}{m_s}} - \sqrt{\dfrac{m_u}{m_c}} \rlp & r \llp \sqrt{ \dfrac{m_{u}}{m_{c}} } \dfrac{ m_s}{m_b} - \ds \sqrt{m_u \over m_c} {m_{c} \over m_t} \rlp \\[8pt]
 \llp \sqrt{\dfrac{m_d}{m_s}}-\sqrt{\dfrac{m_u}{m_c}} \rlp & 1 &  r \llp \dfrac{m_s}{m_b} - \dfrac{m_c}{m_t} \rlp \\[10pt]
 r \llp \ds \sqrt{m_d \over m_{s}} \dfrac{m_c}{m_t} -  \ds \sqrt{m_d \over m_s} {m_{s} \over m_b} \rlp & - r \llp \dfrac{m_s}{m_b} - \dfrac{m_c}{m_t} \rlp & 1 \\
\end{pmatrix} .
\end{align}
Here, we omit the minus sign in the square root ($\sqrt {- m_{u} / m_{c}} \to \sqrt{m_{u} / m_{c}}$).
It predicts $V_{cb}$ and $V_{ts}$ at leading order as follows

\begin{align}
V_{cb} \simeq - V_{ts} \simeq  r \llp \dfrac{m_s}{m_b} - \dfrac{m_c}{m_t} \rlp .
\end{align}
If the parameters $K_{f}, \d_{f}, \e_{f}$ have $CP$ phases, 
each CKM matrix element obtains overall phases and relative phases, such as $\sqrt{\frac{m_d}{m_s}} - \sqrt{\frac{m_u^{}}{m_c}} \to e^{i \phi} \llp \sqrt{\frac{m_d}{m_s}} - e^{i \h} \sqrt{\frac{m_u^{}}{m_c}} \rlp $. 
In particular, the best value of $\chi^{2}$ fit 
$r = \sqrt {81/32} \simeq 1.59$ \cite{Mondragon:1998gy}  
gives excellent agreement between the prediction and the observation
of absolute values of the CKM matrix elements.

\section{$SU(4)_{c} \times SU(2)_{L} \times SU(2)_{R}$ model with democratic texture}

In order to explore a more fundamental understanding of flavor,  
building some unified model is a standard method. 
The grand unified theory (GUT) with the democratic texture is only discussed in \cite{Fukugita:1998kt,Fukugita:1999xb}, as far as the author knows. 
However, since these papers assumed degenerated $Y_{\n}$, 
unification between $Y_{\n}$ and other $Y_{f}$ is difficult.
In this paper, we attempt to build another unified model with the democratic texture, 
which has some unification between $Y_{\n}$ and $Y_{u}$.
Since the $S_{3L} \times S_{3R}$ flavor symmetry is chiral
\footnote{In the $SO(10)$ GUT, the flavor symmetry should be single $S_{3}$, and the condition $c_{f} = 0$ similar to Eq.~(\ref{0jouken}) should be assumed.},
the unified gauge group is assumed to be Pati--Salam (PS) type $SU(4)_{c} \times SU(2)_{L} \times SU(2)_{R}$  ($G_{422}$) \cite{Pati:1974yy}.

To produce realistic fermion masses, 
we consider the minimal contents of Higgs fields with the following representations under the $G_{422}$ group: 
\begin{align}
\Phi : (1,2,2), ~~~~ \S : (15,2,2), ~~~~ \D_{R} : (10,1,3) .
\end{align}
Although other representations are also possible, such as (4,1,2) in \cite{Blazek:2003wz,King:2014iia}, 
we consider only renormalizable interactions to control Yukawa interactions. 

The field contents of the unified model are in Table 2.
These Higgs contents are sufficient to break the PS gauge group $G_{224}$ to the SM gauge group $G_{SM}$.
For example, a breaking scheme of the gauge symmetry with these Higgs contents 
is discussed in the context of the noncommutative geometry \cite{Aydemir:2015nfa,Aydemir:2016xtj}. 
We do not discuss the energy scales and order of the symmetry breakings. 
However, the final result is considered to be rather independent from breaking schemes.
\begin{table}[htb]
  \begin{center}
    \begin{tabular}{|c|ccc|cc|} \hline
           & $SU(4)_{c}$ & $SU(2)_{L}$ & $SU(2)_{R}$ & $S_{3L}$ & $S_{3R}$ \\ \hline \hline
      $\Psi_{Li} = (q_{Li}^{\a}, l_{L i})$ & \bf 4 & \bf 2 & \bf 1 & $\bf 1_L + 2_L$ & $\bf 1_R$    \\
      $\Psi_{Ri} = (q_{Ri}^{\a} , l_{Ri})$ & \bf 4 & \bf 1 & \bf 2 & $\bf 1_L$ &  $\bf 1_R + 2_R$ \\ \hline
      $\Phi$ & \bf 1 & \bf 2 & \bf 2 & $\bf 1_L$ &  $\bf 1_{R}$ \\ 
      $\S$ & \bf 15 & \bf 2 & \bf 2 & $\bf 1_L$ &  $\bf 1_{R}$ \\ 
      $\D_{R}$ & \bf 10 & \bf 1 & \bf 3 & $\bf 1_L$ &  $\bf 1_R $ \\ \hline
    \end{tabular}
    \caption{The charge assignments of the SM fermions and Higgs fields under the gauge and the flavor symmetries.}
  \end{center}
\end{table}

The renormalizable Yukawa interactions invariant under $G_{422}$ are found to be
\begin{align}
\Lg_{\rm Yukawa} = 
\bar \Psi_{R i} (Y^{1}_{ij} \Phi + Y^{15}_{ij} \S) \Psi_{L j} 
+ {\rm H.c.} \, .
\end{align}
Note that Yukawa matrices $Y^{1,15}$ become symmetric matrices if we impose 
the left-right symmetry between $\Psi_{L} \leftrightarrow \Psi_{R}$.
These $Y^{1,15}$ are divided into 
$S_{3L} \times S_{3R}$ preserving and breaking parts respectively:
\begin{align}
Y^{1} = K_{1} D + \d Y_{1}, ~~ 
Y^{15} = K_{15} D + \d Y_{15} .
\end{align}
In order to obtain the desirable masses and mixings,  
we assume $K_{15} = 0$ and $\d Y_{1}$ does not have $S_{3L} \times S_{3R}$ breaking elements $\d_{f}$.  
Then $Y_{15}$ is treated as a perturbation, as in the previous study \cite{Fukugita:1998kt}.
Vacuum expectation values of these Higgs fields are taken to be
\begin{align}
\vev{\Phi} = {\rm Diag} (1,1,1,1) \times \diag{v_{u}^{1}}{v_{d}^{1}} ,  ~~~
 \vev{\S} = {\rm Diag} (1,1,1,-3) \times \diag{v_{u}^{15}}{v_{d}^{15}} , 
\end{align}
in the representation space of $\Psi_{L,R} = (q_{L,R}^{1}, q_{L,R}^{2}, q_{L,R}^{3}, l_{L,R})$.

This setup leads to the following mass matrices \cite{Babu:1992ia, Masiero:2002jn, Bertolini:2006pe}
\begin{align}
M_u &= v_{u}^{1} (K_{1} D + \d Y_{1}) + v_{u}^{15} \delta Y_{15} 
= v_{u}^{1} K_{1} D + v_{u}^{1} \d Y_{1} + v_{u}^{15} \delta Y_{15} , \\
M_\nu^{D} &= v_{u}^{1} (K_{1} D + \d Y_{1}) - 3 v_{u}^{15} \delta Y_{15} 
= v_{u}^{1} K_{1} D + v_{u}^{1} \d Y_{1} -3  v_{u}^{15} \delta Y_{15} , \\
M_d &= v_{d}^{1} (K_{1} D + \d Y_{1}) + v_{d}^{15} \delta Y_{15} 
= v_{d}^{1} K_{1} D + v_{d}^{1} \d Y_{1} + v_{d}^{15} \delta Y_{15} , \\
M_e &= v_{d}^{1} (K_{1} D + \d Y_{1}) - 3 v_{d}^{15} \delta Y_{15} 
= v_{d}^{1} K_{1} D + v_{d}^{1} \d Y_{1} -3  v_{d}^{15} \delta Y_{15} . 
\end{align}
In particular, effective Yukawa matrices are explicitly written as
\begin{align}
Y_{u} &= 
\begin{pmatrix}
0 & \e_{u} & 0 \\
\e_{u} & \d_{u} & r \d_{u} \\
0 & r \d_{u} & K_{u}
\end{pmatrix} , 
~~~ 
Y_{d} = 
\begin{pmatrix}
0 & \e_{d} & 0 \\
\e_{d} & \d_{d} & r \d_{d} \\
0 & r \d_{d} & K_{d}
\end{pmatrix} ,  \\
Y_{\n} &= 
\begin{pmatrix}
0 & \e_{\n} & 0 \\
\e_{\n} & \d_{\n} & r \d_{\n} \\
0 & r \d_{\n} & K_{\n}
\end{pmatrix} , 
~~~ 
Y_{e} = 
\begin{pmatrix}
0 & \e_{e} & 0 \\
\e_{e} & \d_{e} & r \d_{e} \\
0 & r \d_{e} & K_{e}
\end{pmatrix} ,  
\end{align}
with 
\begin{align}
K_{u,d} = K_{\n,e} , ~~  \d_{u,d} = - {1\over 3} \d_{\n, e} , ~~~ \e_{u,d} = \e_{\n, e} .
\end{align}
These conditions lead to the famous Georgi--Jarlskog relation \cite{Georgi:1979df}
\begin{align}
m_{d} = 3m_{e},~~~~~ m_s= {1\over 3}m_{\m},~~~~~ m_b=m_{\t} ,
\end{align}
and similar formulae hold for up-type fermions.

\section{Analysis of the right-handed Majorana neutrino mass matrix}

In this section, we analyze the right-handed neutrino mass matrix $M_{R}$ 
in the PS model with the four-zero Yukawa textures. 
Many papers have studied this kind of model, such as 
SO(10) GUT with the four-zero texture \cite{Nishiura:1999yt,Matsuda:1999yx, Bando:2003wb,Bando:2004hi}. 
However, the purpose of this paper is to analyze texture of $M_{R}$ quantitatively
in a united model with the democratic texture.

$M_{R}$ emerges from the following interaction
\begin{align}
\Lg_{\rm Majorana} = 
\bar \Psi_{R i}^{c} Y^{10}_{ij} \D_{R} \Psi_{R j} + {\rm H.c.} \, ,
\end{align}
when $\D_{R}$ obtain a vacuum expectation value
\begin{align}
\vev{\D_{R}} = {\rm Diag} ( 0,0,0,1) \times 
\begin{pmatrix}
0 & 0 \\ v_{R} & 0
\end{pmatrix} .
\end{align}
Because $Y^{10}$ is transformed as $(\bf 1_{R} + 2_{R}) \times (1_{R} + 2_{R})$, 
it has two $S_{3R}$ invariant terms \cite{Fukugita:1998vn}
\begin{align}
Y^{10} = K_{10} D + c_{10} 1_{3} + \d Y_{10}. 
\end{align}
where $ 1_{3}$ is the $3 \times 3$ identity matrix.

To obtain the observed light neutrino masses, 
we assume the type-I seesaw mechanism \cite{seesaw} 
\begin{align}
m_{\n} = {v^{2} \over 2} Y_{\n}^{T} M_{R}^{-1} Y_{\n} .
\label{seesaw}
\end{align}
In this case, 
\begin{align}
\d Y_{10} \gg c_{10} \simeq 0,
\label{0jouken}
\end{align}
is required by phenomenological reason.
The numerical analysis shown later reveals that $Y_{\n}$ with a large $c_{10} \gg \d Y_{10}$ are incompatible 
to obtain the observed large neutrino mixings. 

If the flavor symmetry breaking $S_{3L} \times S_{3R} \to S_{2L} \times S_{2R} \to 0$ 
also controls the structure of $M_{R}$, and if there is no fine-tuning between the parameters,  
the form of $M_{R}$ should be the following cascade texture in Table 1:
\begin{align}
M_{R} \sim v_{R}
\begin{pmatrix}
\e & \e & \e \\
\e & \d & \d \\
\e & \d & 1 \\
\end{pmatrix} .
\label{cascade}
\end{align}
The light neutrino mass, Eq.~(\ref{seesaw}), 
 is diagonalized by 
\begin{align}
m_{\n} \equiv V_{\n}^{*} m_{\n}^{\rm diag} V_{\n}^{\dg}, 
\end{align}
where $m_{\n}^{\rm diag} = {\rm diag} (m_{1}, m_{2}, m_{3})$.
This mass matrix is rewritten as 
\begin{align}
m_{\n} = B_{e}^{*} U_{\rm PMNS}^{*} \, m_{\n}^{\rm diag} \, U_{\rm PMNS}^{\dg} B_{e}^{\dg}, 
\end{align}
with the neutrino mixing matrix $U_{\rm PMNS} = B_{e}^{\dg} V_{\n}$ and $B_{e}$~(\ref{Bf}) for the charged leptons. 

Ignoring all of the complex phases for simplicity, 
we can reconstruct $M_{R}$ by the seesaw formula:
\begin{align}
M_{R} &= {v^{2} \over 2} Y_{\n}^{T} m_{\n}^{-1} Y_{\n}  \\
&= {v^{2} \over 2} Y_{\n}^{T} B_{e} U_{\rm PMNS} \, (m_{\n}^{\rm diag})^{-1} \, U_{\rm PMNS}^{T} B_{e}^{T}     Y_{\n} .
\end{align}

As a benchmark, $M_{R} (\L_{\rm GUT}) = Y_{\n} (\L_{\rm GUT})^{T} m_{\n} (\L_{\rm GUT}) Y_{\n} (\L_{\rm GUT}) $ at the GUT scale $\L_{\rm GUT} = 2 \times 10^{16} \GeV$ can be evaluated as
\begin{align}
{M_{R} (\L_{\rm GUT})\over [\GeV] } &\simeq 
{[\meV] \over m_{1} }
\begin{pmatrix}
 1.876 \times 10^7 & -3.623 \times 10^8 & -1.009 \times 10^{11} \\
 -3.623 \times 10^8 & 6.996 \times 10^9 & 1.948 \times 10^{12} \\
 -1.009 \times 10^{11} & 1.948 \times 10^{12} & 5.424 \times 10^{14} \\
\end{pmatrix} \label{MRshowm1}
\\&+ {[\meV]\over m_{2}}
\begin{pmatrix}
 3.302 \times 10^7 & -2.173 \times 10^9 & -2.849 \times 10^{11} \\
 -2.173 \times 10^9 & 1.429 \times 10^{11} & 1.874 \times 10^{13} \\
 -2.849 \times 10^{11} & 1.874 \times 10^{13} & 2.457 \times 10^{15} \\
\end{pmatrix} \label{MRshowm2}
\\&+{[\meV]\over m_{3}}
\begin{pmatrix}
 6.255 \times 10^7 & 1.012 \times 10^{10} & 3.975 \times 10^{11} \\
 1.012 \times 10^{10} & 1.637 \times 10^{12} & 6.431 \times 10^{13} \\
 3.975 \times 10^{11} & 6.431 \times 10^{13} & 2.526 \times 10^{15} \\
\end{pmatrix} \label{MRshowm3} .
\end{align}
The parameters used here are summarized in Table.~3. 
\begin{table} [ht]
\centering
\begin{math}
\begin{array}{|c|c||c|c|}
\hline
 m_u\,(\text{MeV}) & 0.48 & \theta_{12}^l  & 33.48^{\circ} \\ 
 m_c\,(\text{GeV}) & 0.235 & \theta_{23}^l & 42.3^{\circ} \\
 m_t\,(\text{GeV}) &74.0 & \theta_{13}^l & 8.5^{\circ} \\
 m_e\,(\text{MeV}) & 0.470 & \D m_{31}^{2} (\eV^{2}) & 2.457 \times 10^{-3} \\ 
 m_\m\,(\text{MeV}) & 99.14 & \D m_{21}^{2} (\eV^{2}) & 7.50 \times 10^{-5} \\
 m_\t \,(\text{MeV}) & 1685 & & \\ \hline
\end{array}
\end{math}
\vspace{0.5cm}
\caption{ Input values (for the SM) at the scale $M_{\rm GUT} = 2 \times 10^{16}$ GeV. Similar parameter set is used in \cite{Altarelli:2013aqa}.}
\label{obs}
\end{table}
The fermion masses at the GUT scale $m_{f} (\L_{\rm GUT})$ are taken from \cite{Xing:2007fb}. 
In most cases of this model, 
the order of light neutrino masses $m_{i}$ becomes the normal hierarchy. 
The inverted hierarchy $m_{1} \simeq m_{2} \gg m_{3}$ 
is unnatural because the hierarchy of $M_{R}$ should overcome the ratio 
$m_{t}^{2} / m_{c}^{2}$. 
The renormalization of the neutrino mass can be neglected 
for the normal hierarchy case \cite{Antusch:2003kp,Antusch:2005gp}. 
Then, neutrino mixing angles and mass square differences are taken from the latest global fit \cite{Gonzalez-Garcia:2015qrr}, without renormalization running. 
A similar parameter set is used in \cite{Altarelli:2013aqa}.

Eqs.~(\ref{MRshowm1})\,-\,(\ref{MRshowm3}) shows that 
the right-handed neutrino mass matrix $M_{R} \sim Y_{u}^{T} Y_{u}$ rather tends to be
 the waterfall texture in Table 1,
\begin{align}
M_{R} \sim v_{R}
\begin{pmatrix}
\e^{2} & \e \d & \e  \\
\e \d  & \d^{2} & \d \\
\e & \d & 1 \\
\end{pmatrix} ,
\label{waterfall}
\end{align}
 for each small mass eigenvalue $m_{i}$. 
Then, it seems to be difficult to explain this texture by 
the breaking scheme $S_{3L} \times S_{3R} \to S_{2L} \times S_{2R} \to 0$.
Hereafter we precisely check the form of the $M_{R}$ by numerical analysis. 

\subsection{Numerical results}

Using the mass difference values $\D m_{3i}^{2}$ in Table 3,
\begin{align}
m_3 = \pm \sqrt{m_1^2+2457} \, [\meV], ~~~ m_2 = \pm \sqrt{m_1^2+75} \, [\meV] ,
\end{align}
the mass matrix $M_{R}$~(\ref{MRshowm1})\,-\,(\ref{MRshowm3}) is expressed as a function of $m_{1}$, $M_{R} (\L_{\rm GUT}) = M_{R} (m_{1})$.

\begin{figure}[h]
\begin{center}
\begin{tabular}{ccc}
   \includegraphics[width=8cm,clip]{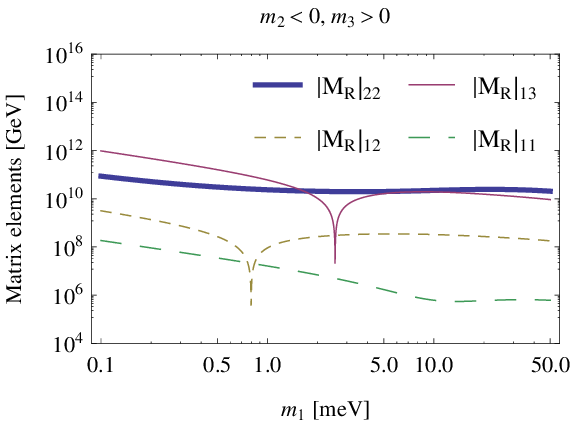} &
      \includegraphics[width=8cm,clip]{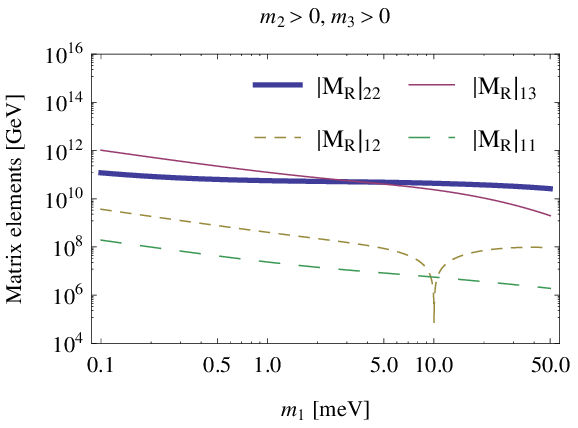} \\
   \includegraphics[width=8cm,clip]{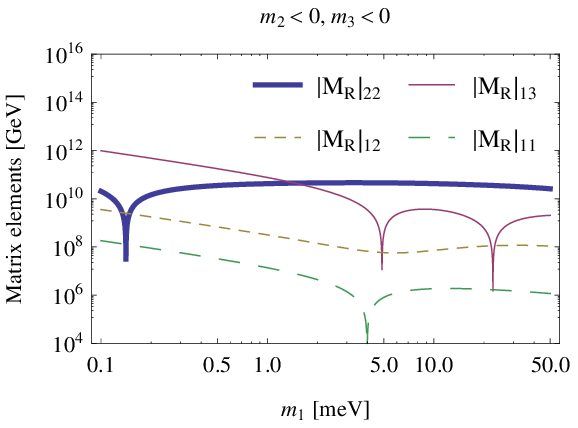} &
      \includegraphics[width=8cm,clip]{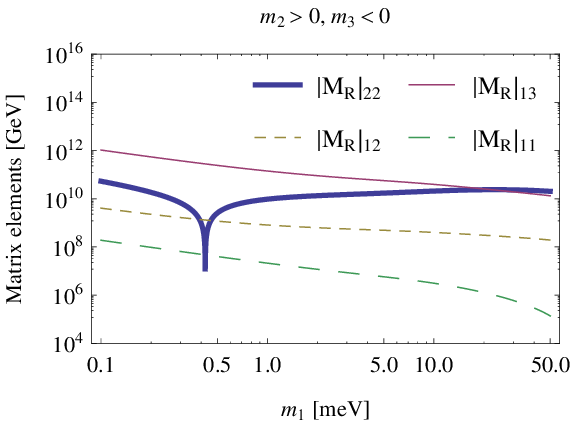} \\
\end{tabular} 
\caption{Lighter matrix elements $(M_{R})_{11}, (M_{R})_{12}, (M_{R})_{13}$, and $(M_{R})_{22}$ of the $M_{R} (m_{1})$ at the GUT scale $\L_{\rm GUT} = 2 \times 10^{16} \, [\GeV]$, as a function of $m_{1}$. The signatures of $m_{2}$ and $m_{3}$ are taken as the top of the figures. }
\end{center}
\end{figure}

Figure 1 shows lighter matrix elements $(M_{R})_{11}, (M_{R})_{12}, (M_{R})_{13}$, and $(M_{R})_{22}$ of the $M_{R} (m_{1})$ at the GUT scale $\L_{\rm GUT} = 2 \times 10^{16} \, [\GeV]$, as a function of $m_{1}$. The signatures of $m_{2}$ and $m_{3}$ are taken as the top of the figures. 
From Fig.~1, we can see the hierarchical structure of the $M_{R}$. 
These matrix elements basically behave like the waterfall texture $(M_{R})_{22} \sim (M_{R})_{13} \gg (M_{R})_{22} \gg  (M_{R})_{11}$. 
Several changes of sign are due to cancellations among Eqs.~(\ref{MRshowm1})\,-\,(\ref{MRshowm3}). 

This behavior shows that 
 the cascade texture $(M_{R})_{22} \gg (M_{R})_{13} \sim (M_{R})_{22} \sim  (M_{R})_{11}$ 
cannot be realized without fine-tunings of parameters in this model. 
In particular, the four-zero texture for $M_{R}$ (equivalent to $(M_{R})_{11} = (M_{R})_{13} = 0$), is also difficult to realize  without fine-tuning. 
However, in this analysis, approximate four-zero texture $(M_{R})_{12} \gg (M_{R})_{13} \sim (M_{R})_{11}$ is realized around $m_{1} \sim 4 \, \meV$ with $m_{2,3} < 0$.

\vspace{12pt}

So far,  the parameters of the model have been assumed to be real. 
Here we will discuss the effect of $CP$ phases shortly. 
\begin{figure}[h]
\begin{center}
\begin{tabular}{ccc}
   \includegraphics[width=8cm,clip]{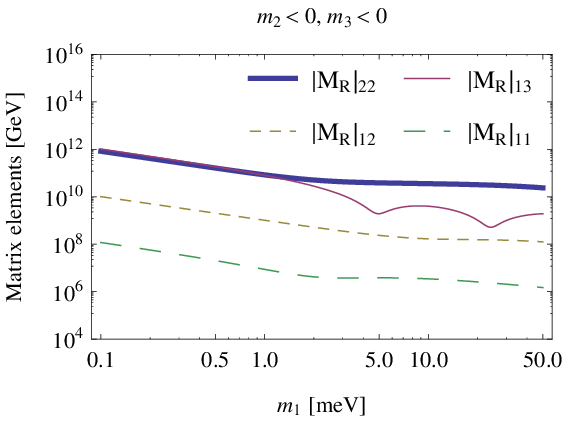} &
   \includegraphics[width=8cm,clip]{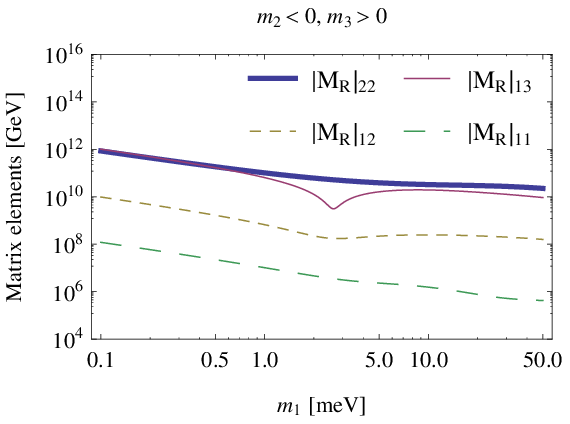} 
\end{tabular} 
\caption{Lighter matrix elements $(M_{R})_{11}, (M_{R})_{12}, (M_{R})_{13}$, and $(M_{R})_{22}$ of the $M_{R} (m_{1})$, with finite dirac $CP$ phase $\d_{CP} = \pi/2$ of the () PMNS matrix. Other parameters are taken to be the same as Fig. 1 (for $m_{2}$, only negative sign $m_{2} < 0$ is presented).}
\end{center}
\end{figure}
Figure 2 shows lighter matrix elements $(M_{R})_{11}, (M_{R})_{12}, (M_{R})_{13}$, and $(M_{R})_{22}$ of the $M_{R} (m_{1})$, with finite dirac $CP$ phase $\d_{CP} = \pi/2$ of the Pontecorvo--Maki--Nakagawa--Sakata (PMNS) matrix. Other parameters are taken to be the same as Fig. 1 (for $m_{2}$, only negative sign $m_{2} < 0$ is presented). 
In Fig.~2, the cancellations of $(M_{R})_{ij}$ found in Fig.~1 vanish by the finite $CP$ phases, 
and the cascade texture is evidently impossible with this parameter set. 
By assuming finite $CP$ phases for other parameters, we found that 
the cancellations are basically smoothed or vanished. 
It is plausible that $M_{R}$ is strongly tend to be the waterfall texture~(\ref{waterfall}). 
Therefore, in this democratic matrix approach, 
a model with type-I seesaw and up-type Yukawa unification $Y_{\n} \simeq Y_{u}$ 
 basically requires fine-tunings between parameters (including its $CP$ phases, errors of the input parameters, and gauge symmetry breaking schemes). 
If we realize the breaking scheme  $S_{3L} \times S_{3R} \to S_{2L} \times S_{2R} \to 0$
by some mechanism, 
the sector of $\n_{R}$ might be too complicated 
to obtain cascade $Y_{f}$ and  waterfall $M_{R}$ in a unified picture. 
Therefore, it seems to be more realistic 
to consider universal waterfall textures for both $Y_{f}$ and $M_{R}$, 
e.g., by the radiative mass generation \cite{Babu:1990fr} or the Froggatt--Nielsen mechanism \cite{Froggatt:1978nt}. 

\subsection{Mass eigenvalues and thermal leptogenesis}

%
\begin{figure}[h]
\begin{center}
\begin{tabular}{ccc}
\includegraphics[width=8cm,clip]{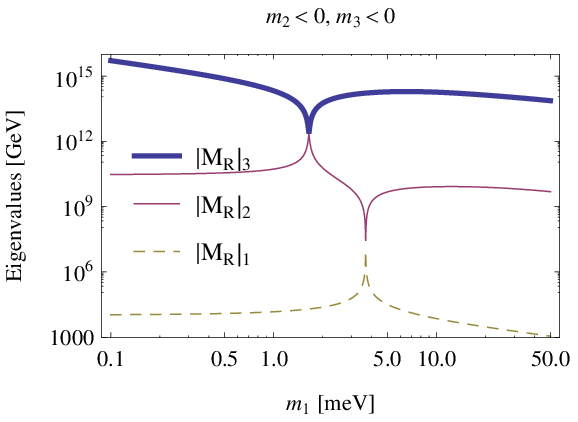} &
\includegraphics[width=8cm,clip]{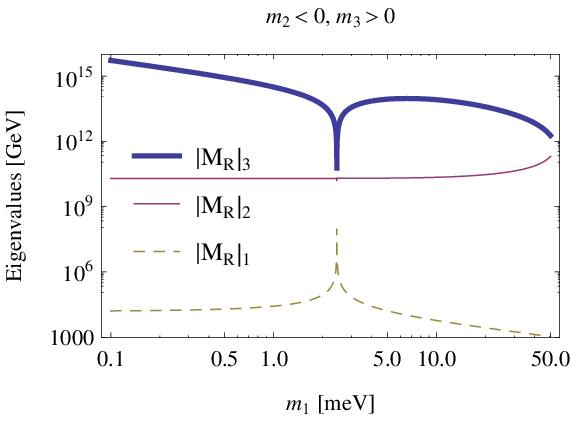} 
\end{tabular} 
\caption{Three mass eigenvalues $M_{Ri}$ of the $M_{R} (m_{1})$ at the GUT scale $\L_{\rm GUT} = 2 \times 10^{16} \, [\GeV]$, as a function of $m_{1}$. The parameters are taken to be the same as Fig. 1 (for $m_{2}$, only negative sign $m_{2} < 0$ is presented). }
\end{center}
\end{figure}

Figure 3 shows three mass eigenvalues $M_{Ri}$ of the $M_{R} (m_{1})$ at the GUT scale $\L_{\rm GUT} = 2 \times 10^{16} \, [\GeV]$, as a function of $m_{1}$. 
The parameters are taken to be the same as Figure 1 (for $m_{2}$, only negative sign $m_{2} < 0$ is presented).
%
Basically the eigenvalues $M_{Ri}$ are strongly hierarchical, 
because $M_{R}$ has large hierarchy such as $M_{R} \sim Y_{u}^{T} Y_{u}$. 
The largest eigenvalue $M_{R3}$ changes its sign around $m_{1} \sim 2 \, \meV$. 
This is due to cancellation for the 33 element of $M_{R}$, between Eq.~(\ref{MRshowm1}) and Eq.~(\ref{MRshowm2}) around the region $m_{2} \sim 5 m_{1}$. 
Similarly, the cancellation for $(M_{R})_{11}$ induces the change of sign for two smaller eigenvalues, $M_{R1}$ and $M_{R2}$. 

These figures exhibit that the lightest mass eigenvalue tends to be rather small $M_{R1} \lesssim 10^{5} \,  \GeV$, except the cancellation regions. 
The successful thermal leptogenesis \cite{Fukugita:1986hr} requires 
$M_{R1} > 4.9 \times 10^{8} \, \GeV$ for the hierarchical $M_{Ri}$  \cite{Buchmuller:2002rq, Giudice:2003jh}.
Then, it is nearly impossible to explain the observed baryon asymmetry by the thermal leptogenesis in this model.   
The resonant leptogenesis \cite{Pilaftsis:1997jf, Pilaftsis:2003gt} would be possible in the cancellation region with  $M_{R1} \simeq M_{R2}$ ($m_{3} < 0, m_{1} \simeq 3 \meV$). 
Similar results for SO(10) are found in Ref.~\cite{Bando:2004hi}. 
However, this cancellation region can be easily vanished by finite CP phases. 
Therefore, successful leptogenesis also requires fine-tunings of the parameters in this model. 

In this study, we assume only renormalizable Yukawa interactions. 
However, this strong tendency to the waterfall texture originates from 
the seesaw relation $M_{R} \sim Y_{u}^{T} Y_{u}$. 
Therefore, it would be rather robust for non-renormalizable Yukawa interactions,
as far as the type-I seesaw mechanism is assumed. 

\section{Conclusions}

In this paper, we attempt to build a unified model with the democratic texture, 
which has some unification between up-type Yukawa interactions $Y_{\n}$ and $Y_{u}$.
Since the $S_{3L} \times S_{3R}$ flavor symmetry is chiral, 
the unified gauge group is assumed to be Pati-Salam (PS) type $SU(4)_{c} \times SU(2)_{L} \times SU(2)_{R}$  ($G_{422}$).
The breaking scheme of the flavor symmetry is considered to be
 $S_{3L} \times S_{3R} \to S_{2L} \times S_{2R} \to 0$. 
In this picture, the four-zero texture is desirable for realistic mass and mixings. 
This texture is realized by a specific representation for the second breaking of the $S_{3L} \times S_{3R}$ flavor symmetry. 

Assuming only renormalizable Yukawa interactions, type-I seesaw mechanism, 
and neglecting $CP$ phases for simplicity, the right-handed neutrino mass matrix 
$M_{R}$ can be reconstructed from low energy input values.
Numerical analysis shows that 
the texture of $M_{R}$ basically behaves  
like the waterfall texture in Table 1. 
Since $M_{R}$ tends to be the cascade texture in the democratic texture approach, 
a model with type-I seesaw and up-type Yukawa unification $Y_{\n} \simeq Y_{u}$ 
 basically requires fine-tunings between parameters (including its $CP$ phases, errors of the input parameters, and schemes of gauge symmetry breaking). 
If we realize the breaking scheme $S_{3L} \times S_{3R} \to S_{2L} \times S_{2R} \to 0$
by some mechanism, 
the sector of $\n_{R}$ might be too complicated 
to obtain cascade $Y_{f}$ and  waterfall $M_{R}$ in a unified picture. 
Therefore, it seems to be more realistic 
to consider universal waterfall textures for both $Y_{f}$ and $M_{R}$, 
e.g., by the radiative mass generation or the Froggatt--Nielsen mechanism. 

Moreover, analysis of eigenvalues shows that 
the lightest mass eigenvalue $M_{R1}$ is too light to account 
the baryon asymmetry of the universe by the thermal leptogenesis.
Although the resonant leptogenesis might be possible, 
it also requires fine-tunings of parameters. 

In this study, we assume only renormalizable Yukawa interactions. 
However, this strong tendency to the waterfall texture originates from 
the seesaw relation $M_{R} \sim Y_{u}^{T} Y_{u}$. 
Therefore, it would be rather robust for non-renormalizable Yukawa interactions,
as far as the type-I seesaw mechanism is assumed. 

\section*{Acknowledgement}

This study is financially supported by the Iwanami Fujukai Foundation, and the Sasakawa Scientific Research Grant from The Japan Science Society, No.~28-214.


\end{document}